\documentclass[12pt]{iopart}
\usepackage[latin1]{inputenc}
\begin{document}

\title{A conjugate for the Bargmann representation}

\author{A. D. Ribeiro$^1$, F. Parisio$^{2,3}$ and M. A. M. de Aguiar$^4$}

\address{$^1$Departamento de F\'{\i}sica, Universidade Federal do
Paraná, 81531-990, Curitiba, PR, Brazil}

\address{$^2$Departamento de F\'{\i}sica, Universidade Federal
de Pernambuco, 50670-901, Recife, PE, Brazil}

\address{$^3$ Instituto de Física, Universidade Federal de Alagoas, 57072-970,
Maceió, Alagoas, Brazil.}

\address{$^4$Instituto de F\'{\i}sica ``Gleb Wataghin'',
Universidade Estadual de Campinas, Unicamp 13083-970, Campinas,
S\~ao Paulo, Brazil}

\begin{abstract}

In the Bargmann representation of quantum mechanics, physical
states are mapped into entire functions of a complex variable
$z^*$, whereas the creation and annihilation operators
$\hat{a}^\dagger$ and $\hat{a}$ play the role of multiplication and
differentiation with respect to $z^*$, respectively. In this paper
we propose an alternative representation of quantum states,
conjugate to the Bargmann representation, where the roles of
$\hat{a}^\dagger$ and $\hat{a}$ are reversed, much like the roles
of the position and momentum operators in their respective
representations. We derive expressions for the inner product that
maintain the usual notion of distance between states in the Hilbert
space. Applications to simple systems and to the calculation of
semiclassical propagators are presented.

\end{abstract}

\section{Introduction}

In quantum mechanics, the position and the momentum of a particle
are represented by operators $\hat{q}$ and $\hat{p}$ satisfying the
canonical commutation relation $[\hat{q},\hat{p}]=i\hbar$. The
eigenstates of $\hat{q}$, obeying $\hat{q}|q\rangle = q |q\rangle$,
form a complete set and define the {\it coordinate representation},
where state kets $|\psi\rangle$ are mapped into square-integrable
wavefunctions $\psi(q)=\langle q|\psi\rangle$ with
$\hat{q}|\psi\rangle \rightarrow \langle q|\hat{q}|\psi\rangle =
q\psi(q)$ and $\hat{p}|\psi\rangle \rightarrow \langle
q|\hat{p}|\psi\rangle = -i\hbar\partial \psi(q)/\partial q$.
Similarly, the eigenstates of $\hat{p}$ define the {\it momentum
representation}, where $|\psi\rangle \rightarrow
\tilde\psi(p)=\langle p|\psi\rangle$ with $\hat{p}|\psi\rangle
\rightarrow \langle p|\hat{p}|\psi\rangle=p\tilde\psi(p)$ and
$\hat{q}|\psi\rangle \rightarrow \langle
p|\hat{q}|\psi\rangle=i\hbar\partial \tilde\psi(p)/\partial p$. The
two representations are said to be conjugate to each other and are
related by the Fourier transformation
\begin{equation}
\psi(q) = \int \langle q|p\rangle \tilde\psi(p) dp =
\frac{1}{\sqrt{2\pi\hbar}}\int \tilde\psi(p) e^{ipq/\hbar} dp.
\label{e1}
\end{equation}

The interplay between the position and the momentum representations
is of great importance in the quantum theory. Although the
information contained in either representation is the same, the
clarity and simplicity of a calculation depend strongly on which
representation is chosen. Simple illustrations in one-dimension are
the time independent Schrödinger equation for the square barrier
potential, which is very simple in the coordinate representation,
and the linear potential $V(q)=q$, which can be solved immediately
in the momentum representation. This is particularly useful to
calculate WKB wavefunctions near turning points, where the
coordinate representation is singular.

More elaborate applications involve semiclassical approximations
for time dependent problems, such as the propagator
$K(q_f,q_i,T)=\langle q_f|e^{-i\hat{H}T/\hbar}|q_i \rangle$, which
depends on classical trajectories starting at $q_i$ and ending at
$q_f$ after a time $T$. The semiclassical approximation for $K$
diverges at the so called focal points, where $\partial
p_i/\partial q_f \rightarrow \infty$, and is inaccurate in a whole
vicinity of these points \cite{berrymount}. In general, a focal
point in the position representation is not a simultaneous focal
point in the momentum representation and, as proposed by Maslov
\cite{maslov,maslov2}, one can switch between the two
representations to pass by the focal point. In other words, the
semiclassical approximation for $K(p_f,q_i,T)=\langle
p_f|e^{-i\hat{H}T/\hbar}|q_i \rangle$ is well behaved when
calculated at the same trajectory where $K(q_f,q_i,T)$ is
divergent, and can be Fourier transformed to produce accurate
results for the original propagator $K(q_f,q_i,T)$.

Besides the position and momentum representations, a different set
of continuous basis states can be defined with the help of coherent
states, whose importance in physics has been recognized since the
early days of quantum mechanics
\cite{bargmann,glauber,Klau78,Klau85,perelomov,gilmore}. In the
special case of the harmonic oscillator, coherent states are
closely associated with the creation and annihilation operators
\begin{equation}
  \label{e2}
  \hat{a} = \frac{1}{\sqrt{2}}\left( \frac{\hat{q}}{b}+i
              \,\frac{\hat{p}}{c} \right), \qquad
  \hat{a}^\dagger = \frac{1}{\sqrt{2}}\left( \frac{\hat{q}}{b}-i
              \,\frac{\hat{p}}{c} \right),
\end{equation}
where $b=\sqrt{\hbar/(m\omega)}$ and $c=\sqrt{m\hbar\omega}$, with
$m$ and $\omega$ the mass and frequency of the oscillator,
respectively. The commutation relation
$[\hat{a},\hat{a}^{\dagger}]=1$, together with the eigenvalue
equation $\hat{a}|z\rangle = z|z\rangle$, define an alternative
representation of quantum mechanics which was introduced by Fock
and studied in detail by Bargmann \cite{bargmann}, who lent his
name to the theory, Glauber \cite{glauber} and others
\cite{Klau78,Klau85,perelomov,gilmore}. In the {\it Bargmann}
representation, the state $|\psi\rangle$ is mapped into an entire
function $\psi(z^*)=\langle z|\psi\rangle$ of the complex variable
\begin{equation}
 \label{e3}
 z^* =  \frac{1}{\sqrt{2}}\left( \frac{q}{b}-i\,\frac{p}{c} \right),
\end{equation}
where $\hat{a}^{\dagger}|\psi\rangle \rightarrow \langle
z|\hat{a}^{\dagger}|\psi\rangle = z^*\psi(z^*)$ and
$\hat{a}|\psi\rangle \rightarrow \langle
z|\hat{a}|\psi\rangle=\partial \psi(z^*)/\partial z^*$. The
(unnormalized) Bargmann states $|z\rangle$ are given by
\begin{equation}
  \label{e4}
  |z \rangle = e^{z \, \hat{a}^{\dagger}} |0 \rangle ,
\end{equation}
where $|0 \rangle$ is the ground state of the harmonic oscillator.
These states are related to the (normalized) canonical coherent
states $|z\rangle\rangle$ by
$|z\rangle\rangle=e^{-|z|^2/2}|z\rangle$. The real numbers $q$ and
$p$ are the average values of the position and momentum operators
in the state $|z\rangle\rangle$.

Contrary to the position and the momentum representations, the
Bargmann representation lacks a dual counterpart. Indeed, since the
operator $\hat{a}^{\dagger}$ does not have eigenstates, it is not
possible to map $|\psi\rangle$ into $\psi(z)$ such that
$\hat{a}|\psi\rangle$ is mapped into $z\psi(z)$. It might be argued
that, because it is a phase-space representation, where both $q$
and $p$ participate simultaneously, a conjugate representation is
simply not needed. It was initially thought, for instance, that the
phase-space propagator $K(z^*_f,z_i,T)=\langle
z_f|e^{-i\hat{H}T/\hbar}|z_i \rangle$ would be free of focal points
\cite{mcdonald,klauder1,leboeuf,voros}. Focal points, however, do
exist in the coherent state propagator
\cite{Adachi,Klau95,Tan98,rib04} and in mixed representations as
well \cite{Hel87,Hel88,Shu95,Shu96,Agu05} and the application of
the Maslov method would require a conjugate representation for the
Bargmann states.

The existence of phase space focal points motivated the definition
of an application that could play the role of a conjugate
representation for the Bargmann states \cite{ribprl} and that was
successfully used in applications of the Maslov method
\cite{rib08,rib08a}. Its relation to the Bargmann representation,
however, is not as simple as the relation between the position and
momentum representations, but it does comply with the basic
requirements of a dual map. The purpose of this paper is to
formalize this conjugate representation and to study it in more
detail.

The paper is organized as follows: in section 2 we review some of
the main ingredients of the Bargmann representation. In section 3
we define its conjugate counterpart in terms of line integrals in
the complex plane and study some of its properties. In section 4 we
present alternative formulas where the line integrals are replaced
by integrals over the entire complex plane and in section 5 we show
a few simple applications. Finally, in section 6, we summarize our
results.

\section{The Bargmann representation}

In the Bargmann formalism, a state ket $|\psi\rangle$ is
represented in phase space by its projection onto a non-normalized
coherent state
\begin{equation}
\psi(z^*)=\langle z|\psi\rangle\ = \langle 0|e^{z^* a}|\psi\rangle\;.
\end{equation}
The state of the system is completely determined by the entire
function $\psi(z^*)$. The resolution of unit is expressed in terms
of the integral
\begin{equation}
\hat{I}=\int \frac{{\rm d}^2 z~}{\pi} e^{-|z|^2}\, |z \rangle \langle z|
\equiv \int\frac{dq dp}{2\pi\hbar}\; e^{-|z|^2}\, |z \rangle \langle z|
\equiv  \int {\rm d}^2\mu(z)~ |z \rangle \langle z|\;,
\label{unit}
\end{equation}
so that the inner product between $| \psi\rangle$ and $|
\phi\rangle$ reads
\begin{equation}
\langle \psi| \phi \rangle=\int {\rm d}^2\mu(z)~
\psi^*(z^*) \phi(z^*) \equiv (\psi,\phi),
\label{inner}
\end{equation}
where the last equality also defines the inner product between the
two corresponding entire functions.

The Bargmann space ${\cal F}$ is composed of the entire functions
$\psi(z^*)$ such that $(\psi,\psi) < \infty$. The mapping between
$\psi(q)$ and $\psi(z^*)$ can be constructed explicitly as
\begin{equation}
\psi(z^*) = \int dq \langle z|q\rangle \langle q|\psi\rangle\;
= \pi^{-1/4}b^{-1/2}\int dq ~ e^{-\frac{1}{2}({z^*}^2+q^2/b^2)+
\sqrt{2}z^*q/b}~ \psi(q),
\end{equation}
with its inverse given by
\begin{equation}
\psi(q) = \pi^{-1/4}b^{-1/2}\int {\rm d} \mu(z) ~
e^{-\frac{1}{2}(z^2+q^2/b^2)+\sqrt{2}zq/b}~ \psi(z^*).
\end{equation}

If $\psi$ and $\phi$
are expressed as a power series as
\begin{equation}
\psi(z^*) = \sum_{n=0}^{\infty} a_n z^{*n}/\sqrt{n!} \quad, \quad \qquad
\phi(z^*) = \sum_{n=0}^{\infty} b_n z^{*n}/\sqrt{n!}
\label{series}
\end{equation}
the overlap reduces to
\begin{equation}
\langle \psi| \phi \rangle=\sum_{n=0}^{\infty} a_n^* b_n.
\end{equation}
Therefore, the set of functions
\begin{equation}
\phi_n(z^*)= \langle z|n\rangle = z^{*n}/\sqrt{n!}
\label{phin}
\end{equation}
forms a complete orthonormal set in ${\cal F}$, where $|n\rangle$
are the eigenstates of the underlying harmonic oscillator,
Eqs.~(\ref{e2})--(\ref{e4}).

In the Bargmann representation it is convenient to express
observables in terms creation and annihilation operators. The
action of these operators on $| \psi\rangle$ yields
\begin{equation}
\langle z| \hat{a}^{\dagger} | \psi\rangle=z^*\psi(z^*)\;, \qquad \quad
\langle z| \hat{a} | \psi\rangle=\frac{\partial}{\partial z^*}\psi(z^*)\;.
\end{equation}
Any observable $\hat{A}(\hat{a}^{\dagger},\hat{a})$ is, therefore,
written in the Bargmann representation as
$\hat{A}_B=\hat{A}(z^*,\frac{\partial}{\partial z^*})$. This
identification is valid for any ordering of the operators since the
commutation relation $[\hat{a},\hat{a}^{\dagger}]=1$ is preserved,
i. e., $\left[\frac{\partial}{\partial z^*},z^*\right]=1$. Thus,
one can recast the time-independent Schroedinger equation $\hat{H}
| \psi\rangle=E | \psi\rangle $ as $\hat{H}_B \psi(z^*)=E \psi(z^*)$.
For the simple harmonic oscillator $\hat{H}_B=\hbar
\omega(z^*\frac{\partial}{\partial z^*}+1/2)$ and we get
\begin{equation}
\hbar \omega\left(z^*\frac{\partial}{\partial z^*}+\frac{1}{2}\right)
u_n(z^*)= E_n u_n(z^*)\;,
\end{equation}
whose solutions are exactly the normalized functions $\phi_n(z^*)$,
defined by Eq.~(\ref{phin}), with eigenvalues $E_n=\hbar
\omega(n+1/2)$, $n=0,1,2,\cdots$.

Before closing this section we derive Bargmann's reproducing kernel
from the resolution of unit~(\ref{unit}). Multiplying this equation
on the right by $|\psi\rangle$ and on left by $\langle w|$ we
obtain
\begin{equation}
\psi(w^*)=\int {\rm d}^2 \mu(z)~ \langle w|z \rangle \psi(z^*) =
\int {\rm d}^2 \mu(z)~ e^{w^*z} \psi(z^*).
\label{repker}
\end{equation}
The reproducing kernel ${\cal K}(w^*,z)= \langle w|z\rangle =
e^{w^*z}$ plays the role of the delta function in the position and
momentum representations and will be important to derive some
useful relations in the next sections.

\section{The Conjugate Application}

\subsection{Basic definitions}
\label{suba}

Let $|\psi\rangle$ be a state ket and $\psi(z^*) = \langle z|\psi
\rangle$ its Bargmann representation. For each coherent state
$|w\rangle$ we define the application $|\psi\rangle \longrightarrow
f_\psi (w)$ by
\begin{equation}
{f}_\psi (w) = \int_\gamma
\frac{\langle z|\psi \rangle}
{\langle z|w \rangle}~dz^* =
\int_\gamma \psi(z^*)~ e^{-z^* w} dz^*
\label{dualdef}
\end{equation}
and its inverse by
\begin{equation}
\psi(z^*) = \frac{1}{2\pi i}\int_{\gamma'} {f}_\psi(w)
\langle z|w \rangle~dw = \frac{1}{2\pi i} \int_{\gamma'}
{f}_\psi(w)~e^{z^* w} dw.
\label{dualdefI}
\end{equation}
The integration paths $\gamma$ and $\gamma'$ will be defined below.

Although the denominator in Eq.~(\ref{dualdef}) might look unusual,
it is really a direct generalization of the transformation between
the coordinate and the momentum representations, which can be
written as
\begin{displaymath}
\tilde\psi(p)= \frac{1}{2\pi\hbar} \int \frac{\langle q|\psi
\rangle}{\langle q|p \rangle} ~ dq \qquad{\rm and} \qquad \psi(q) =
\int  \langle q| p \rangle \langle p| \psi \rangle dp. \label{FT}
\end{displaymath}
However, while both $\psi(q)$ and $\tilde\psi(p)$ are matrix
elements between the ket $|\psi\rangle$ and a bra, $f_\psi(w)$ is
not itself a matrix element. Moreover, the application is linear in
$|\psi\rangle$, since
\begin{equation}
{f}_{\alpha\psi+\beta\phi} (w) =\alpha f_\psi + \beta f_\phi,
\end{equation}
but not in $|w\rangle$. For this reason the nomenclature {\it
conjugate application} is preferred instead of conjugate
representation.

\subsection{Action of operators}
\label{subb}

Before we specify the integration paths $\gamma$ and $\gamma'$ we
explore the action of operators on the conjugate functions.
Consider two states $|\psi_1\rangle = \hat{a}^\dagger|\psi\rangle$
and $|\psi_2\rangle = \hat{a}|\psi\rangle$, whose Bargmann
representations are given, respectively, by $\psi_1 (z^*) =z^*
\psi(z^*)$ and $\psi_2 (z^*) = \frac{\partial\psi(z^*)}{\partial
z^*}$. The corresponding conjugate functions are, according
to~(\ref{dualdef}),
\begin{equation}
f_{\psi_1}(w) =
\int_{\gamma} z^* \psi(z^*) e^{-z^*w}dz^* =
-\frac{\partial}{\partial w}f_{\psi}(w)
\label{prop1}
\end{equation}
and
\begin{equation}
f_{\psi_2}(w) =
\int_{\gamma}\frac{\partial\psi(z^*)}{\partial z^*}
e^{-z^*w}dz^*  = wf_{\psi}(w),
\label{prop2}
\end{equation}
where we have integrated by parts and assumed that
$\psi(z^*)e^{-z^*w}$ vanishes at the extremes of $\gamma$ (see
comment after Eq.(\ref{anI}) in next subsection). Consequently, if
$|\phi\rangle = \hat A (\hat a, \hat a^\dagger) |\psi\rangle$ and
$\phi(z^*)= \hat{A}_B\left(\frac{\partial}{\partial
z^*},z^*\right)\psi(z^*)$, then
\begin{equation}
f_{\phi}(w) = \hat{A} \left(w,-\frac{\partial}{\partial w}\right)
f_{\psi} (w) \equiv \hat{A}_C f_{\psi} (w)
\end{equation}
since the commutation relation $[\hat{a},\hat{a}^{\dagger}]=1$ is
preserved in the form $[w,-\partial/\partial w]=1$.

The duality between the two representations is therefore expressed
by the action of $\hat{a}$ and $\hat{a}^\dagger$ on the
corresponding functions $\psi(z^*)$ and $f_\psi(w)$:
\begin{equation}
\begin{array}{l}
\hat{a} \qquad \overrightarrow{{\scriptstyle Bargmann}} \qquad \frac{\partial}{\partial z^*}
\qquad \overrightarrow{{\scriptstyle Conjugate}} \qquad w  \\ \\
\hat{a}^\dagger \qquad \overrightarrow{{\scriptstyle Bargmann}} \qquad z^*
\qquad \overrightarrow{{\scriptstyle Conjugate}} \qquad -\frac{\partial}{\partial w} .
\end{array}
\end{equation}
In particular, the Schrödinger Equation in the space of functions
$f_{\psi}(w)$ becomes
\begin{equation}
\hat{H}_C\left(- \frac{\partial}{\partial w},w\right) f_\psi(w) =
E f_\psi(w).
\label{Sch}
\end{equation}
For the harmonic oscillator we obtain
\begin{equation}
\hbar\omega \left(- \frac{\partial}{\partial w} w +
\frac{1}{2}\right) f_\psi(w) = E f_\psi(w)
\label{Schoh}
\end{equation}
and the eigenfunctions and eigenvalues can be immediately
calculated as
\begin{equation}
f_n(w) = \frac{\sqrt{n!}}{w^{n+1}}~, \qquad \qquad E_n=\hbar\omega(n+1/2),
\label{eeho}
\end{equation}
where the choice of normalization is justified in the next subsection.

\subsection{Integration paths}

In order to define the paths $\gamma$ and $\gamma'$ in
Eqs.~(\ref{dualdef}) and (\ref{dualdefI}), we consider the
expansion of a general ket $|\psi\rangle$ in the harmonic
oscillator basis $\{|n\rangle\}$, namely,
$|\psi\rangle=\sum_{n=0}^{\infty}a_n |n\rangle$. The Bargmann
representation of $|\psi\rangle$ is
\begin{equation}
\psi(z^*) = \sum_{n=0}^{\infty} a_n \langle z|n \rangle =
\sum_{n=0}^{\infty} a_n \phi_n(z^*) =
\sum_{n=0}^{\infty} \frac{a_n {z^*}^n}{\sqrt{n!}} ,
\label{psiexp}
\end{equation}
where
\begin{equation}
a_n = \frac{1}{\sqrt{n!}}\int \psi(z^*)z^n \, {\rm d}^2\mu(z).
\label{an}
\end{equation}
Inserting~(\ref{psiexp}) in Eq.~(\ref{dualdef}), we find
\begin{equation}
f_{\psi}(w) = \sum_{n=0}^{\infty} a_n f_{\phi_n}(w),
\end{equation}
where
\begin{equation}
f_{\phi_n}(w) = \int_\gamma \phi_n(z^*) e^{-z^*w}dz^* =
\frac{1}{\sqrt{n!}} \int_\gamma {z^*}^n e^{-z^*w}dz^*.
\end{equation}

We now demand that $f_{\phi_n}(w)=f_n(w)$, given by
Eq.~(\ref{eeho}). This is achieved by converting the line integral
into a Laplace transform. Writing $z$ and $w$ in terms of polar
variables, $z = r_z e^{i\theta_z}$ and $w =r_w e^{i\theta_w}$, the
exponent of the integrand becomes $- z^*w = - r_z r_w
e^{i(\theta_w-\theta_z)}$. The path $\gamma$ is fixed by choosing
$\theta_z = \theta_w$ and $r_z$ going from 0 to $\infty$. In fact,
since the function being integrated is analytic, it suffices to
take paths that can be deformed into this one. Explicitly we obtain
\begin{equation}
f_{\phi_n}(w) =  \frac{e^{-i(n+1)\theta_w}}{\sqrt{n!}}
\int_{0}^{\infty} r_z^n e^{-r_zr_w} dr_z
= \frac{w^{-(n+1)}}{\sqrt{n!}} \Gamma(n+1) =
\frac{\sqrt{n!}}{w^{n+1}},
\label{psiexpIa}
\end{equation}
which leads to the Laurent series
\begin{equation}
f_{\psi}(w) = \sum_{n=0}^{\infty} a_n \frac{\sqrt{n!}}{w^{n+1}}.
\label{psiexpI}
\end{equation}
Provided the sum on the right side converges we can also write
\begin{equation}
a_n = \frac{1}{\sqrt{n!}}\int  f_{\psi}(w) w^{n+1}\, {\rm d}^2\mu(w).
\label{anI}
\end{equation}
This choice of $\gamma$ also guarantees the correctness of
Eq.(\ref{prop2}) for functions that can be expressed as power
series like~(\ref{psiexp}), since $\partial \psi/\partial z^*$ does
not depend on $a_0$.

Alternatively, using this integration path directly into
Eq.~(\ref{dualdef}) leads to
\begin{equation}
f_{\psi}(w) = \int_0^\infty \psi(r_z e^{-i\theta_w})
e^{-r_z r_w-i\theta_w} dr_z =
\frac{1}{w}\int_0^\infty \psi\left(\frac{x}{w}\right) e^{-x} dx .
\label{dualdef2}
\end{equation}

Similarly, the inverse transform of $f_{\phi_n}(w)$ can be written as
\begin{equation}
\frac{1}{2\pi i} \int_{\gamma'} f_{\phi_n} (w) e^{wz^*} dw
= \frac{\sqrt{n!}}{2\pi i} \int_{\gamma'} \frac{e^{wz^*}}{w^{n+1}} dw,
\end{equation}
which can be performed by using Cauchy's residue theorem, when
$\gamma'$ is conveniently chosen and the integral becomes a Mellin
integral. Since the pole is located at the origin, $\gamma'$ should
be perpendicular to the straight line connecting the origin with
$z$, crossing the real axis on the positive (negative) side if
$Re(z^*)>0$ ($Re(z^*)<0$). Then, we get
\begin{eqnarray}
\frac{1}{2\pi i} \int_{\gamma'} f_{\phi_n} (w) e^{wz^*} dw =
\frac1{\sqrt{n!}}\left.\left(\frac{d^n  e^{wz^*}}{dw^n}\right)\right|_{w=0}
= \frac{{z^*}^n}{\sqrt{n!}} = \phi_n(z^*). \label{op1}
\end{eqnarray}
Therefore, the general inverse formula can be written as
\begin{equation}
\psi(z^*) = \frac{1}{2\pi z^*} \int_{-\infty-i\epsilon}^{\infty-i\epsilon}
f_\psi\left(\frac{iv}{z^*}\right) e^{iv} dv,
\label{dualdef2I}
\end{equation}
where $\epsilon$ is a positive number.

Alternative expressions for the mappings between $\psi(z^*)$ and
$f_\psi(w)$ that avoid the line integrals will be given in section
\ref{subc}.

\subsection{Scalar product}

The scalar product between two kets $|\psi\rangle$ and
$|\phi\rangle$ can be obtained starting from
\begin{equation}
\langle \psi|\phi\rangle =
\int \psi^*(z^*) \phi(z^*)~{\rm d}^2 \mu(z)
= \sum_{n,m} a_m^* b_n \int \phi^*_m(z^*) \phi_m(z^*){\rm d}^2 \mu(z),
\end{equation}
where $|\psi\rangle=\sum_{n=0}^\infty a_n|n\rangle$ and
$|\phi\rangle=\sum_{n=0}^\infty b_n|n\rangle$. Using
Eqs.~(\ref{op1}) and~(\ref{dualdef2I}) we obtain
\begin{equation}
\langle \psi|\phi\rangle =
\frac{1}{4\pi^2}\sum_{n,m=0}^{\infty} a_m^* b_n\sqrt{m!n!}{\cal A}_{mn},
\label{eqpi1}
\end{equation}
where
\begin{equation}
{\cal A}_{mn} = \int
\left[{z^*}^m\int_{-\infty}^{\infty}
\frac{e^{\epsilon+iy}}{(\epsilon+iy)^{m+1}}dy \right]^*
\left[{z^*}^n\int_{-\infty}^{\infty}
\frac{e^{\epsilon+iy}}{(\epsilon+iy)^{n+1}}dy  \right]
{\rm d}^2 \mu(z).
\label{Amn1}
\end{equation}
The integration over ${\rm d}^2 \mu(z)$ gives
\begin{equation}
\int {z}^{n}{z^*}^{m} ~{\rm d}^2 \mu(z) =
m! \delta_{mn}.
\end{equation}
In addition, the integral inside the first brackets of
Eq.~(\ref{Amn1}) can be evaluated by residues resulting in
$2\pi/m!$. Thus,
\begin{equation}
\begin{array}{lll}
{\cal A}_{mn} &=& 2\pi \delta_{mn}
\int_{-\infty}^{\infty}
\frac{e^{\epsilon+iy}}{(\epsilon+iy)^{n+1}}dy \\
&=&2\pi \delta_{mn}
\int_{0}^{\infty} \left[
\frac{e^{\epsilon+iy}}{(\epsilon+iy)^{n+1}} +
\frac{e^{\epsilon-iy}}{(\epsilon-iy)^{n+1}}\right] dy
\end{array}
\label{Amn2}
\end{equation}
and Eq.~(\ref{eqpi1}) becomes
\begin{equation}
\begin{array}{lll}
\langle \psi|\phi\rangle &=&
\frac{1}{2\pi}\sum_{n,m=0}^{\infty} a_m^* b_n\sqrt{m!n!}
\delta_{mn}
\int_{0}^{\infty} \left[
\frac{e^{\epsilon+iy}}{(\epsilon+iy)^{n+1}} +
\frac{e^{\epsilon-iy}}{(\epsilon-iy)^{n+1}}\right] dy.
\end{array}
\label{eqpi2}
\end{equation}
At last, we use the identity
\begin{equation}
\frac{\delta_{mn}e^{\epsilon\pm iy}}{(\epsilon\pm iy)^{n+1}} =
\frac{1}{2\pi} \int_{0}^{2\pi} \frac{e^{i(m-n)\theta} e^{\epsilon \pm
iy}d\theta}{(\epsilon\pm iy)^{\frac{n+1}{2}} (\epsilon\pm
iy)^{\frac{m+1}{2}}}
\end{equation}
and change the variable of integration in Eq.~(\ref{eqpi2}) as
$y=r^2$ obtaining
\begin{equation}
\langle \psi|\phi\rangle =
\frac{1}{2\pi^2}
\int_{0}^{\infty} r dr \int_{0}^{2\pi}d\theta~ {\rm F}_\epsilon(r,\theta),
\label{eqpi3}
\end{equation}
where
\begin{equation}
\begin{array}{lll}
{\rm F}_\epsilon(r,\theta) &=&
f_{\psi}^*(w_{(+)}e^{-i\pi/4})f_{\phi}(w_{(-)}e^{+i\pi/4})e^{+i(r^2-i\epsilon)}\\
&+&
f_{\psi}^*(w_{(-)}e^{+i\pi/4})f_{\phi}(w_{(+)}e^{-i\pi/4})e^{-i(r^2+i\epsilon)}
\end{array}
\end{equation}
and $w_{(\pm)} = \sqrt{r^2\pm i\epsilon}~e^{i\theta}$. Finally,
taking the limit $\epsilon\rightarrow0$,
$w_{(+)}=w_{(-)}=re^{i\theta}$ and, defining $w_1=w
e^{-i\frac{\pi}{4}}$ and $w_2=w e^{i\frac{\pi}{4}}$ we obtain
\begin{equation}
\begin{array}{lll}
\langle \psi|\phi\rangle &=&
\frac{1}{\pi^2} \int \left[
f^*_{\psi}( w_1)
f_{\phi}(w_2) e^{i|w|^2}
+
f^*_{\psi}(w_2)
f_{\phi}(w_1) e^{-i|w|^2}
\right]{\rm d}^2 w .
\end{array}
\end{equation}

\section{Alternative formal transformations}
\label{subc}

\subsection{The coherent state}

The Weyl displacement operator $\hat{D}=e^{z\hat{a}^\dagger}$ in
the dual space becomes $\hat{D}_C = e^{-z\frac{\partial}{\partial
w}}$. As a consequence, since $|z\rangle = e^{z\hat{a}^{\dagger}}
|0\rangle = \hat{D} |0\rangle$ and the conjugate of the ground
state is $f_0(w) = \frac{1}{w}$, we find
\begin{equation}
f_z(w) = e^{-z\frac{\partial}{\partial w}}
\left(\frac{1}{w}\right)
=\frac{1}{w} + \frac{z}{w^2}+\frac{z^2}{w^3} + \cdots =
\frac{1}{w-z}
\label{disp}
\end{equation}
for $|z/w|<1$. This shows that $\hat{D}_C$ also acts as a
displacement operator in the dual space. Although the series does
not converge inside the circle $|z/w|=1$, we analytically extend it
to $f_z(w)=1/(w-z)$ to the whole complex plane, except for $w=z$.
This continuation is justified because the path of integration
$\gamma'$ in the inverse transformation (\ref{dualdefI}) can always
be chosen to lie outside the circle $|z/w|=1$ and, therefore, the
integral is independent of $f_z(w)$ in this region. Note that the
result~(\ref{disp}) can also be obtained using the basic definition
(\ref{dualdef2}). This time the convergence region is for $w$
outside the circle of radius $|z|/2$ centered on $z/2$, which is
less restrictive than that obtained via displacement operator.
Analytic continuation is then similar to that done for the Laplace
transform of the exponential function.

Equation (\ref{disp}) provides an important formal expression of
$f_{\psi}(w)$. Starting from the expansion for $|\psi\rangle$ in
coherent states
\begin{equation}
|\psi\rangle = \int \langle z|\psi\rangle |z\rangle \,{\rm d} ^2 \mu(z)
\label{exppsi}
\end{equation}
and transforming both sides we obtain
\begin{equation}
f_{\psi}(w) = \int \psi(z^*) f_z(w) \,{\rm d} ^2 \mu(z) .
\label{dualdefws}
\end{equation}
The integral in equation~(\ref{dualdefws}) is over the whole
phase-space, avoiding the cumbersome line integrals of the original
definition. However, this expression is only formal, since going
from  (\ref{exppsi}) to (\ref{dualdefws}) involves the ilegal
interchange of the line integral coming from the definition of
$f_\psi$ and the integral over the complex plane from
(\ref{exppsi}). Nevertheless, expanding $|z\rangle$ in
(\ref{exppsi}) in the harmonic oscillator basis states and doing
the integral term by term we obtain the convergent expression
\begin{equation}
f_{\psi}(w) = \sum_{n=0}^{\infty} \frac{1}{w^{n+1}}
\int z^n \psi(z^*) \,{\rm d} ^2 \mu(z) .
\label{dualdefwss}
\end{equation}
This procedure is equivalent to treating $f_z(w)$ formally as the
series given by (\ref{disp}) and interchange the summation and
integration.

Although the direct transformation~(\ref{dualdefws}) is only formal
and essentially useless, it is possible to write down the inverse
transformation in the same footing which is valid for all
$f_\psi(w)$. Using the formal expression (\ref{dualdefws})
temporarily we write
\begin{equation}
\psi(z^*) = \int A(z^*,t) f_{\psi}(t) ~{\rm d}^2 \mu(t) =
\int A(z^*,t) \psi(w^*) f_w(t) ~{\rm d}^2 \mu(w) ~{\rm d}^2 \mu(t).
\end{equation}
Comparing with the reproducing kernel Eq.(\ref{repker}) we find that
\begin{equation}
\int A(z^*,t) f_w(t)~{\rm d}^2 \mu(t) = e^{z^*w}.
\end{equation}
Going back to the series representation for $f_w(t)$ and exchanging
the summation and integration we recast this equation as
\begin{equation}
\sum_{n=0}^{\infty} w^n \left[ \int \frac{A(z^*,t)}{t^{n+1}}~{\rm d}^2
\mu(t) \right] = e^{z^*w},
\end{equation}
which shows that
\begin{equation}
\int \frac{A(z^*,t)}{t^{n+1}}~{\rm d}^2 \mu(t) \equiv \frac{{z^*}^n}{n!}.
\end{equation}
Differentiating both sides $n$ times with respect to $z^*$ gives
one. Comparing again with Eq.(\ref{repker}) we find that $A(z^*,t)
= t \exp{\{z^*t\}}$ and
\begin{equation}
\psi(z^*) = \int w \,e^{z^*w} f_{\psi}(w) \,{\rm d} ^2 \mu(w).
\label{dualdefwsI}
\end{equation}
Contrary to~(\ref{dualdefws}), this equation is well defined for
all $f_{\psi}(w)$. This is an interesting expression that allows
the construction of matrix elements, such as propagators, from
usual phase space integration of their dual forms. We show in the
appendix how to perform the integral for the basic cases
$|\psi\rangle = |n\rangle$ and $|\psi\rangle = |z\rangle$.

\subsection{Reproducing kernel}

When Eq.(\ref{dualdefwsI}) is substituted back into
(\ref{dualdefws}) we get
\begin{equation}
\begin{array}{ll}
f_{\psi}(w) &= \int w' \,e^{z^*w'} f_{\psi}(w') f_{z}(w) \,{\rm d} ^2 \mu(w')
\,{\rm d} ^2 \mu(z). \\ \\
 & \equiv \int {\cal K}_C(w,w') f_{\psi}(w') \,{\rm d} ^2 \mu(w') .
 \end{array}
\end{equation}

Comparing with (\ref{repker}) we find
\begin{equation}
{\cal K}_C(w,w') = \int w' \,e^{z^*w'}  f_{z}(w) \,{\rm d} ^2 \mu(z)
= w' f_{w'}(w) = \frac{w'}{w-w'}.
\end{equation}
Once again these expressions are only formal and the operational
reproducing equation is
\begin{equation}
f_{\psi}(w) = \sum_{n=0}^{\infty} \frac{1}{w^{n+1}}
\int {w'\,}^{n+1} f_{\psi}(w') {\rm d} ^2 \mu(w').
\end{equation}
%

\subsection{Scalar product}

An expression for the scalar product can be obtained from
Eqs.~(\ref{inner}) and~(\ref{dualdefwsI}):
\begin{equation}
\begin{array}{ll}
(\psi,\phi)&=\int {\rm d}^2 \mu(z)\, \psi^*(z^*) \phi(z^*) \\
&= \int t^*w\, e^{zt^*+z^*w} f_{\psi}^*(t) f_{\phi}(w)
{\rm d}^2 \mu(t)\, {\rm d}^2 \mu(w)\, {\rm d}^2 \mu(z).
\end{array}
\end{equation}
Using (\ref{repker}) again we find
\begin{equation}
\int e^{zt^*+z^*w}\, {\rm d}^2 \mu(z) = e^{t^*w}
\end{equation}
and
\begin{equation}
(\psi,\phi) = \int t^*w\, e^{t^*w} f_{\psi}^*(t) f_{\phi}(w)
{\rm d}^2 \mu(t)\, {\rm d}^2 \mu(w).
\label{prodsc}
\end{equation}

We can check the correctness of this expression by expanding the
exponential in power series and rewriting this as
\begin{equation}
\begin{array}{ll}
(\psi,\phi) &= \sum_{n=0}^{\infty}
\left[ \int \frac{t^{n+1}f_{\psi}(t)}{\sqrt{n!}} {\rm d}^2 \mu(t)\right]^*
\left[ \int \frac{w^{n+1}f_{\phi}(w)}{\sqrt{n!}} {\rm d}^2 \mu(w)\right] \\  \\
&= \sum_{n=0}^{\infty} a_n^* b_n ,
\end{array}
\end{equation}
where we used (\ref{anI}) with $a_n$ and $b_n$ as coefficients for
$\psi$ and $\phi$ respectively (\ref{series}). A mixed
representation for the scalar product can also be obtained by
combining Eqs.~(\ref{prodsc}) and (\ref{dualdefwsI}):
\begin{equation}
(\psi,\phi) = \int t^*\, f_{\psi}^*(t) \phi(t^*) {\rm d}^2 \mu(t).
\label{prodscmix}
\end{equation}
This expression might be useful, considering that for the
eigenstates of the harmonic oscillator $t^*\, f_{\phi_n}^*(t)
\phi_n(t^*)=1$.

\section{Simple examples}

\subsection{The propagator of the harmonic oscillator}

In the Bargmann representation the propagator of the harmonic oscillator is
\cite{ribprl}
\begin{equation}
k(z^*,z_0,t)\equiv \langle z | e^{-i\hat{H}t/\hbar} |
z_0\rangle =  e^{z_0(t) z^* -i \omega t/2},
\label{exact}
\end{equation}
where $z_0(t)=z_0e^{-i\omega t}$. Its conjugate representation
becomes
\begin{equation}
f_{k}(w,z_0,t) = e^{-i \omega t/2} f_{z_0(t)}(w) = \frac{e^{-i \omega t/2}}{w-z_0(t)}.
\label{ktilosc}
\end{equation}
The diagonal conjugate representation becomes simply
\begin{equation}
f_{k}(w,w,t) = \frac{1}{w} \sum_{n=0}^{\infty} e^{-i \omega (n+1/2)t},
\end{equation}
which corresponds directly to its decomposition in eigenfunctions.

\subsection{Position and momentum eigenstates}

Although $\langle z|q\rangle$ does not belong to the Bargmann space
${\cal F}$ of square integrable functions, we can readily write it
down as
\begin{equation}
\langle z|q\rangle = \sum_{n=0}^{\infty} \langle z|n\rangle \langle n|q\rangle =
\sum_{n=0}^{\infty} \frac{{z^*}^n}{\sqrt{n!}} ~\phi_n(q)
\label{zq}
\end{equation}
where
\begin{equation}
\phi_n(q) = \frac{\pi^{-1/4} b^{-1/2}}{2^{n/2} \sqrt{n!}} ~e^{-q^2/2b^2}H_n(q/b),
\label{herm}
\end{equation}
$b=\sqrt{\hbar/m\omega}$ and $H_n$ are the Hermite polynomials.
When (\ref{herm}) is placed into (\ref{zq}) the sum can be
performed an results in the well known expression
\begin{equation}
\langle z|q\rangle =\pi^{-1/4} b^{-1/2}
\exp{\{-\frac{q^2}{2b^2}-\frac{{z^*}^2}{2}+\frac{\sqrt{2}z^*q}{b}\}}.
\label{zq1}
\end{equation}

The expression for $|q\rangle$ in the conjugate representation can
be obtained directly from Eqs.~(\ref{herm}) and (\ref{zq}) and
results in
\begin{equation}
f_q(w) = \pi^{-1/4} b^{-1/2} e^{-q^2/2b^2} \sum_{n=0}^{\infty}
\frac{1}{w^{n+1}} ~2^{-n/2}~H_n(q/b).
\label{fqws}
\end{equation}
Alternatively, using the integral form given by
Eq.~(\ref{dualdef2I}) we find, for $Re(w^2)>0$,
\begin{equation}
f_q(w) = \frac{\pi^{1/4} b^{-1/2}}{\sqrt{2}}
e^{q^2/2b^2-\sqrt{2}wq/b+w^2/2} F(w/\sqrt{2}-q/b),
\end{equation}
where
\begin{equation}
F(u) = \frac{u}{\sqrt{u^2}}
\left[1-{\rm Erf}\left( \frac{u^2}{\sqrt{u^2}}\right)\right]
\end{equation}
and ${\rm Erf}$ is the error function. We have used the notation
$\sqrt{u^2}=r\exp{[i\arctan{(2\theta)}/2]}$ for
$u=r\exp{(i\theta)}$, which is simply $|u|$ if $u$ is real. It can
be shown, using an integral representation for the Hermite
polynomials, that the sum in Eq.~(\ref{fqws}) above can also be
cast in this form for $Re(w^2)>0$. Similar expressions for $f_p(w)$
can be obtained from $f_q(w)$ by replacing $q$ by $p$ and $b$ by
$c=\sqrt{m\hbar\omega}$.

\subsection{Calculation of matrix elements}

Consider the matrix elements
\begin{equation}
\langle z | \hat{X} | z'\rangle =
\langle z| \left[ \frac{b}{\sqrt2}\left( \hat{a}+\hat{a}^\dagger\right)
\right] | z'\rangle
\end{equation}
and
\begin{equation}
\langle z | \hat{P} | z'\rangle = \langle z| \left[
\frac{c}{i\sqrt2}\left( \hat{a}-\hat{a}^\dagger\right) \right] |
z'\rangle.
\end{equation}

The transformed functions become
\begin{equation}
f_{\hat{X} | z'\rangle}(w) =
\frac{b}{\sqrt2}\left( w - \frac{\partial}{\partial w} \right)
f_{z'}(w)
\end{equation}
and
\begin{equation}
f_{\hat{P} | z'\rangle}(w) =
\frac{c}{i\sqrt2}\left( w + \frac{\partial}{\partial w} \right),
f_{z'}(w)
\end{equation}
where $f_{z'}(w)=1/(w-z')$.

\subsection{Semiclassical limit}

In the semiclassical limit, the propagator $k(z_f^*,z_i,T)= \langle
z_f|e^{-i\hat{H}T/\hbar}| z_i \rangle$ can be written in terms of
complex classical trajectories satisfying Hamilton's equations and
certain special boundary conditions. Because the trajectories
involved are complex, $z(t)$ and $z^*(t)$ are independent classical
variables and it is convenient to rename them as $u(t)$ and $v(t)$
respectively. The boundary conditions satisfied by the trajectories
contributing to the semiclassical propagator are then given by
$u(0)=z_i$, $v(T)=z_f^*$. Using the Weyl symbol ${\cal H}$ of the
Hamiltonian operator $\hat{H}$ to govern the classical dynamics,
the semiclassical approximation for $k$ reads
\cite{coelho06,eva07,rib08a}
\begin{equation}
k_{sc}(z_f^*,z_i,T)=
\sum_{\mathrm{traj.}}
\sqrt{\frac{1}{M_{vv}}}~
\exp{\left\{\frac{i}{\hbar}~S \right\}}
\label{sp}
\end{equation}
where $S$ is the action and $M_{vv}$ is an element of the tangent
matrix, that propagates small displacements from the trajectory,
defined by
\begin{eqnarray}
\left(\begin{array}{c}\delta{u}(T)\\\delta{v}(T)\\\end{array}\right) =
\left(\begin{array}{cc}
M_{uu}&M_{uv}\\M_{vu}&M_{vv}\\
\end{array}\right)
\left(\begin{array}{c}\delta {u}(0)\\\delta {v}(0)\\\end{array}\right) \, .
\label{mmatrix}
\end{eqnarray}
The action satisfies the relations
\begin{equation}
\label{mb9}
\frac{\partial S}{\partial z_f^*}= -i\hbar u(T),\qquad
\frac{\partial S}{\partial z_i}= -i\hbar v(0),\qquad
\frac{\partial S}{\partial t}= -{\cal H}(u(T),z_f^*,t)~.
\end{equation}

The conjugate representation of $k_{sc}$ is given, for each
contributing trajectory, by
\begin{equation}
\begin{array}{ll}
\tilde k_{sc} ( w ,z_i,T )& =
\int_{\tilde{C}} k_{sc}(z_f^*,z_i,T) e^{-z_f^*w} dz_f^*  \\ \\
 &=\int_{\tilde{C}} \sqrt{\frac{1}{M_{vv}}}~
\exp{\left\{\frac{i}{\hbar}~(S +i \hbar z_f^*w)\right\}} dz_f^*.
\label{ktil}
\end{array}
\end{equation}
When the integral is performed by the saddle point approximation,
the saddle point condition is given by
\begin{equation}
\frac{\partial S}{\partial z_f^*}= -i\hbar w
\label{spc}
\end{equation}
and the exponent of the transformed expression becomes
\begin{equation}
\tilde{S}(w,z_i,T) = S(z_f,z_i,T) + i\hbar w z_f^*,
\label{na}
\end{equation}
where $z_f^*$ is obtained as a function of $z_i$, $w$ and $T$ from
(\ref{spc}). Equations (\ref{spc}) and (\ref{na}) define a Laplace
transformation and comparison with (\ref{mb9}) reveals that the
trajectory contributing to $\tilde{k}_{sc}$ satisfies $u(0)=z_i$
and $u(T)=w$. When the exponent is expanded to second order around
the saddle point and the resulting quadratic integral is performed,
the conjugate propagator becomes \cite{rib08a}
\begin{eqnarray}
\tilde{k}_{sc}(w, z_i, T ) = \sum_{\mathrm{traj.}}
\sqrt{\frac{1}{M _{u v}}}~
\exp\left\{\frac{i}{\hbar} \tilde{S}(w, z_i,T )
\right\}.
\label{ktil2}
\end{eqnarray}
The whole conjugation process becomes totally analogous to the
conjugation between position and momentum representations. We refer
to Ref.\cite{rib08a} for the details and for applications related
to focal points and the Maslov method.

\section{Summary and Discussion}

The conjugate representation introduced in \cite{ribprl} and
studied here in more detail is not standard. The reason for this
unconventional approach is that, contrary to the annihilation
operator $\hat{a}$, the creation operator $\hat{a}^\dagger$ does
not have eigenstates. However, we have shown that it is still
possible to map Bargmann's entire functions $\psi(z^*)=\langle
z|\psi \rangle$ into a conjugate set of singular functions
$f_{\psi}(w)$ where the roles of $\hat{a}$ and $\hat{a}^\dagger$
are reversed. The map takes the basis functions
$\phi_n={z^*}^n/\sqrt{n!}$ into $f_n=\sqrt{n!}/w^{n+1}$ and
a general entire function $\psi(z^*) = \sum_n a_n
{z^*}^n/\sqrt{n!}$ into $f_{\psi}(w)=\sum_n a_n \sqrt{n!}/w^{n+1}$.

The conjugate mapping is originally defined by means of a contour
integration over a curve $\gamma$ on the $z^*$ complex plane. The
curve is chosen so that $\phi_n(z^*)$ is mapped into $f_n(w)$.
However, when applied to a coherent state $|z_0\rangle$, the
corresponding integral converges to $1/(w-z_0)$ only if $|w/z_0|
>1$ and the conjugate $f_{z_0}(w)$ has to be analytically continued
to the interior of this circle. This continuation has no
consequences for the inversion formula, since the integration curve
$\gamma'$ can be chosen to lie outside this region.

We have shown that other formal transformation formulas can be
derived which avoid the need of contour integrations, replacing
them by integrals over the whole complex plane. These alternative
representations, however, are very sensitive to the limited
convergence of the line integral defining $f_z(w)$, since they make
direct use of this formula. The direct transformation turns out to
be only formal, but the inverse transformation
formula~(\ref{dualdefws}) is well defined and operational.

\begin{appendix}
\section{The phase space inversion formula}

In this appendix we show how the phase space inversion formula
Eq.~(\ref{dualdefwsI}) works for the simple cases where
$|\psi\rangle = |n\rangle$ and $|\psi\rangle = |z_0\rangle$. The
equation is
\begin{equation}
\psi(z^*) = \int w \,e^{z^*w} f_{\psi}(w) \,{\rm d} ^2 \mu(w).
\end{equation}
For $|\psi\rangle = |n\rangle$, $f_\psi(w)=\sqrt{n!}/w^{n+1}$ and
\begin{equation}
\psi(z^*) = \sqrt{n!} \int \frac{1}{w^n} \,e^{z^*w} \,{\rm d} ^2 \mu(w).
\end{equation}
Differentiating with respect to $z^*$ we get
\begin{equation}
\frac{d^n \psi}{d{z^*}^n} = \sqrt{n!} \int \,e^{z^*w} \,{\rm d} ^2 \mu(w) = \sqrt{n!}
\end{equation}
and, therefore, $\psi(z^*) = {z^*}^n /\sqrt{n!}$, which is the
correct result.

For $|\psi\rangle = |z_0\rangle$ we have $f_\psi(w)=1/(w-z_0)$ and
\begin{equation}
\begin{array}{ll}
\psi(z^*) &= \displaystyle{\int \frac{w}{w-z_0} \,e^{z^*w} \,{\rm d} ^2 \mu(w)}
= \displaystyle{\int \left(1+ \frac{z_0}{w-z_0}\right) \,e^{z^*w} \,{\rm d} ^2 \mu(w)} \\ \\
&= \displaystyle{1+z_0 e^{z^*z_0} \int\frac{e^{z^*(w-z_0)}}{w-z_0}
\,{\rm d} ^2 \mu(w)\equiv 1+z_0 e^{z^*z_0} J}.
\end{array} \label{eqa}
\end{equation}
Since $J$ is an analytic function of $z^*$,
\begin{equation}
\frac{d J}{dz^*} = \int \,e^{z^*(w-z_0)} \,{\rm d} ^2 \mu(w) = e^{-z^*z_0}.
\label{ap1}
\end{equation}
To integrate this equation back we must be careful with the
integration constant. For $z_0=0$, $dJ/dz^*=1$ and $J=z^*$, which
is the correct result for the ground state $|0\rangle$. The direct
integration of (\ref{ap1}), on the other hand, gives
$J=-e^{-z^*z_0}/z_0$, which does not satisfy the proper condition
at $z_0=0$. In order to get the correct integration constant we
write
\begin{equation}
\frac{d J}{dz^*} = \sum_{n=0}^\infty \frac{(-1)^n z_0^n {z^*}^n}{n!}
\end{equation}
and
\begin{equation}
J \displaystyle{= \sum_{n=0}^\infty \frac{(-1)^n z_0^n {z^*}^{n+1}}{(n+1)!}
=-\frac{1}{z_0} \sum_{n=1}^\infty \frac{(-z_0^n z^*)^{n}}{n!}
=-\frac{1}{z_0}\left(e^{-z^*z_0}-1\right)}.
\end{equation}
Substituting back into (\ref{eqa}) we obtain the correct result
$\psi(z^*)=e^{z^*z_0}$.

\end{appendix}

\section*{Acknowledgments}

It is a pleasure to thank Hajo Leschke and Alfredo M.O. de Almeida
for interesting discussions. MAMA and ADR acknowledge financial
support from CNPq, FAPESP and FINEP. ADR especially acknowledges
FAPESP for the fellowship $\#$ 04/04614-4. FP thanks financial
support from FACEPE (DCR 0029-1.05/06 and APQ 0800-1.05/06).\\

\newpage
\noindent REFERENCES \\


\begin{thebibliography}{20}

\bibitem{berrymount} Berry~M~V and Mount~K~E 1972 {\it Rep. Prog.
    Phys.} {\bf 35} 315

\bibitem{maslov} Maslov~V~P and Feodoriuk~M~V 1981 {\it
    Semi-Classical Approximations in Quantum Mechanics} (Boston:
    Reidel)

\bibitem{maslov2} Maslov~V~P 1972 {\it Th\'eorie des Perturbations
    et M\'ethodes Asymptotiques} (Paris: Dunod)

\bibitem{bargmann} Bargmann~V 1961 {\it Comm. on Pure and Appl.
    Math.} {\bf 14} 187

\bibitem{glauber} Glauber~R 1963 {\it Phys. Rev.} {\bf 131} 2766

\bibitem{Klau78}  Klauder~J~R 1978 {\it Continuous Representations
    and Path Integrals, Revisited}, in G.~J. {Papa\-do\-pou\-los}
    and J.~T. Devreese, editors, {\em Path Integrals}, NATO
    Advanced Study Institute, Series B: Physics (New York: Plenum)

\bibitem{Klau85} Klauder~J~R and Skagerstam~B~S 1985 {\it Coherent
    States, Applications in Physics and Mathematical Physics}
    (Singapore: World Scientific)

\bibitem{perelomov} Perelomov~A 1986 {\it Generalized Coherent
    States and their Applications} (Berlin: Springer-Verlag)

\bibitem{gilmore} Zhang~W, Feng~D~H and Gimore~R 1990 {\it Rev.
    Mod. Phys.} {\bf 62} 867

\bibitem{mcdonald} Mcdonald~S~W 1985 {\it Phys. Rev. Lett.} {\bf
    54} 1221

\bibitem{klauder1} Klauder~J~R 1986 {\it Phys. Rev. Lett.}  {\bf
    56} 897

\bibitem{leboeuf} Kurchan~J, Leboeuf~P and Saraceno~M 1989 {\it
    Phys. Rev.} A {\bf 40} 6800

\bibitem{voros} Voros~A 1989 {\it Phys. Rev.} A {\bf 40} 6814

\bibitem{Adachi} Adachi~S 1989 {\it Ann. of Phys.} (NY) {\bf 195}
    45

\bibitem{Klau95} Rubin~A and Klauder~J~R 1995 {\it Ann. of Phys.}
    (NY) {\bf 241} 212

\bibitem{Tan98} Tanaka~A 1998 {\it Phys. Rev. Lett.} {\bf 80} 1414

\bibitem{rib04} Ribeiro~A~D, de Aguiar~M~A~M and Baranger~M 2004
    {\it Phys. Rev.} E {\bf 69} 066204

\bibitem{Hel87} Huber~D and Heller~E~J 1987 {\it J. Chem. Phys.}
    {\bf 87} 5302

\bibitem{Hel88} Huber~D, Heller~E~J and Littlejohn~R~G 1988 {\it J.
    Chem. Phys.} {\bf 89} 2003

\bibitem{Shu95} Shudo~A and Ikeda~K~S 1995 {\it Phys. Rev. Lett.}
    {\bf 74} 682

\bibitem{Shu96} Shudo~A and Ikeda~K~S 1996 {\it Phys. Rev. Lett.}
    {\bf 76} 4151

\bibitem{Agu05} de Aguiar~M~A~M, Baranger~M, Jaubert~L, Parisio~F
    and Ribeiro~A~D 2005 {\it J. Phys.} A {\bf 38} 4645

\bibitem{ribprl} Ribeiro~A~D, Novaes~M and de Aguiar~M~A~M 2005
    {\it Phys. Rev. Lett.} {\bf 95} 050405

\bibitem{rib08} Ribeiro~A~D and de Aguiar~M~A~M 2008 {\it Ann.
    Phys.} (NY) {\bf 323} 654

\bibitem{rib08a} Ribeiro~A~D and de Aguiar~M~A~M 2008 {\it J. Phys.
    Conf. Series} {\bf 99} 012016

\bibitem{coelho06} dos Santos~L~C and de Aguiar~M~A~M 2006 {\it J.
    Phys.} A {\bf 39} 13465

\bibitem{eva07} Martín-Fierro~E and Llorente~J~M~G 2007 {\it J.
    Phys.} A {\bf 40} 1065


\end{thebibliography}
\end{document}